\documentclass[a4,12pt]{article}
\newcommand{\be}{\begin{equation}}
\newcommand{\ee}{\end{equation}}
\newcommand{\nonu}{\nonumber}
\begin{document}
\thispagestyle{empty}
\begin{flushright}
ETH-TH/96-15\\
July 1996
\end{flushright}

\bigskip\bigskip\begin{center}
\large{\bf On the Deformation \\ of Time Harmonic 
Flows\footnote{Talk presented at the Centro Stefano
Franscini, Ascona, June 17, 1996.}}\\
\vskip 1.0truecm
{\bf Jens Hoppe\footnote{Heisenberg Fellow, On leave of absence
from Karlsruhe University}} \\
\vskip 2.0truecm
\large{Institut f\"ur Theoretische Physik \\
ETH H\"onggerberg\\ CH-8093 Z\"urich, Switzerland}
\vskip3cm

{\bf Abstract}\\
\end{center}
It is shown that time-harmonic motions of spherical and toroidal
surfaces can be deformed non-locally without loosing the existence
of infinitely many constants of the motion.

\newpage

As pointed out some time ago \cite{Hoppe}, surfaces moving
through ${\bf R}^3$ in such a way that their (normal) velocity
is always equal to the local surface-area density, $\sqrt g$
(divided by some non-dynamical `reference-density', $\rho$),
have the property that the time, at which the surface $\Sigma_t$
reaches a point in space, is a harmonic function. These motions
are related to certain reductions of self-dual $SU(N)$ Yang-%
Mills theories (which play a central role in the construction of
monopoles, cp. \cite{Hitchin}), and the Lax-pair formulation of
the latter can be taken over identically (\cite{Floratos},\cite{Ward}).
One can also show \cite{Bordemann} that time-harmonic flows are, what
remains when projecting certain diffeomorphism invariant Hamiltonian
field theories on to the (integrable) Diff-singlet sector.

In this letter, I would like to show that time-harmonic flows may be
deformed from the `$w_\infty$'-invariant parametrized form to
`$W_\infty$'-invariant motions of parametrized surfaces, while
keeping integrability in the form of infinitely many constants
of motion. Again, the Lax-pair can be taken from the self-dual
Yang-Mills theory, or rather: the Nahm equations (i.e. reduced
self-dual $SU(N)$ Yang-Mills equations) are just a special case
of the most natural first order (in time), quadratically non-linear,
evolution equation for a set of (3) operators, $X_i$.
The specification (of the space of operators) that will correspond
to deformed time-harmonic flows of surfaces of spherical (toroidal)
topology, resp. *-products on $S^2(T^2)$, is discussed in detail.

Let ${\cal A}$ be a non-commutative, associative Algebra (`of
operators') and ${X_i(t)}^N_{i=1}$ a set of timedependent
operators satisfying the non-linear evolution equation(s)
\begin{equation}
\dot{X}_i=\epsilon_{ii_{1}...i_{M}} X_{i_{1}},X_{i_{2}}\dots X_{i_{M}}
\label{eq1}
\end{equation}
$i=1,2,\dots, N=M+1$.

Using \cite{Hoppe6}/\cite{Hoppe7}, one observes that (\ref{eq1})
can be written in the form
\begin{equation}
\dot{L}=\biggl\lbrack L,M_1,\dots, M_{N-2}\biggl\rbrack
\label{eq2},
\end{equation}
where $L$ and the $M$'s $(\epsilon {\cal A}$) are linear
combinations of the $X_i$, depending on $[{N\over 2}]$
spectral parameters, and $[\cdot,\dots,\cdot]$
(a fully antisymmetric map from ${\cal A} \times\dots\times
{\cal A}$ to ${\cal A}$) denotes the natural $M$-commutator,
\begin{equation}
[A_1,\dots,A_M]:=\epsilon^{r_1\dots r_M}
A_{r_{1}} \cdot\dots\cdot A_{r_{M}}
\label{eq3}.
\end{equation}

In particular, one may take
\begin{eqnarray}
L   &= & \mu(X_1+iX_2)-\Bigl ({X_1-iX_2\over \mu}\Bigl )+2X_3 \nonumber\\
M_1 &= & i(\mu(X_1+iX_2)+X_3)
\label{eq4}
\end{eqnarray}
(just as for the reduced self-dual Yang-Mills equations,
see e.g. \cite{Hitchin})
for $N=3$, and (just as in \cite{Hoppe6}/\cite{Hoppe7})
\begin{eqnarray}
L  & = &\mu(X_1+iX_2)+\tilde\mu(X_3+iX_4) \nonu\\
   & - &{1\over\mu}(X_1-iX_2) - {1\over\tilde\mu}(X_3-iX_4)+{\sqrt 8} X_5 \nonu\\
M_1& = &{i\over{\sqrt 2}} \biggl (\mu(X_1+iX_2) + 
       {X_5\over\sqrt 2}\biggl ) \nonu\\
M_2& = &{i\over{\sqrt 2}} \biggl ( \tilde\mu(X_3+iX_4)+
       {X_5\over\sqrt 2}\biggl ) \nonu\\
M_3& = &{-1\over\sqrt 2} \biggl ({(X_1-iX_2)\over\mu} -
       {(X_3-iX_4)\over\tilde\mu}\biggl )
\label{eq5}
\end{eqnarray}
for $N=5$.

If there exists a trace on ${\cal A}$, satisfying
\begin{equation}
Tr AB = Tr BA
\label{eq6},
\end{equation}

\begin{equation}
Q_n : = TrL^n, \quad n\epsilon{\bf N}
\label{eq7}
\end{equation}
will be automatically time-independent only for $N=3$;
for odd $N>3$, at least $Q_1$ and $Q_2$ are conserved,
while for even $N$ not even the basic $M$-commutator
is traceless.

In the following, I will restrict myself to the case $N=3$, i.e.
\begin{equation}
\dot{X}_i = {1\over 2}\epsilon_{ijk} [X_j, X_k]
\label{eq8}
\end{equation}
which, if the $X_i$ were finite-dimensional matrices, are just
`Nahm's equations'. They still trivially `are', for infinite
matrices with finitely many non-zero coefficients, but `all'
other infinite dimensional choices for ${\cal A}$ (or rather:
an infinite-dimensional Lie-algebra, ${\cal L}$) are of quite different
nature, and it seems that only the time-harmonic \cite{Hoppe}/%
\cite{Bordemann}, $w_\infty$-invariant case (area-preserving limit
of $SU(N)$ \cite{Hoppe8},\cite{Floratos},\cite{Ward}), where ${\cal L}$
is the Lie algebra of (non-constant) symplectic diffeomorphisms
of $S^2$ or $T^2\dots$, (\ref{eq8}) becoming the following set of first
order partial differential equations for time-dependent functions
on a two-dimensional manifold $\sum$,
\begin{equation}
\dot x_i={1\over 2} \epsilon_{ijk} {\epsilon^{rs}\over \rho(\phi)}
{\partial x^j\over\partial\phi^r}{\partial x^k\over\partial\phi^s}
\label{eq9},
\end{equation}
has previously been considered in the literature (\cite{Floratos}
- - \cite{Hoppe7}). Here, I would like to consider *-product
deformations of (\ref{eq9}), which amounts to (for
$S^2$, \cite{Bordemann9}) choosing ${\cal A}$ to be the
enveloping algebra of $SO(3)$, divided by the `Casimir-ideal',
or (for $T^2$) specific subclasses of infinite-dimensional
matrices with only finitely many non-zero off-diagonals
(cp. \cite{Floratos10}). Both series of infinite dimensional
`$W_\infty$'-algebras admit an invariant trace, making
(\ref{eq7}) time-independent for all $n$. 

Let me first discuss
the `spherical type' $W^\infty$-algebras (cp. \cite{Bordemann9},
\cite{Hoppe11}).

Let ${\cal G}$ be a semi-simple Lie-algebra,
$\bigl\{T_a\bigr\}^{d=dim{\cal G}}_{a=1}$
a basis of ${\cal G}$,
\begin{equation}
[T_a, T_b] = f^c_{ab} T_c \quad\quad abc=1\dots d
\label{eq10}
\end{equation}
and $U(\cal G)$ be the associative algebra (over {\bf C})
of polynomials
\begin{equation}
T=c_T\cdot{\bf 1} + \sum c^{a_{1}\dots a_{2}} T_{a_{1}}\dots T_{a_{l}}
\label{eq11},
\end{equation}
modulo (\ref{eq10}) (i.e. the universal enveloping algebra). The
center of $U$ is generated by $r= {\rm rank} \cal G$
`Casimirs' $C_1, \dots, C_r$, and $U$ may be divided by the
sum of the $r$ two-sided ideals
\begin{equation}
I_j=(C_j-\lambda_j{\bf 1})U, \quad\quad \lambda_j\epsilon {\bf C}
\label{eq12};
\end{equation}
resulting in $U_{\lambda=(\lambda_{1},\dots,\lambda_{r}})
({\cal G}$), the algebra of polynomials
\begin{equation}
T^{(\lambda)} = c_T{\bf 1} +\sum c_T^{a_{1}\dots a_{l}}
T^{(\lambda)}_{a_{1}}\cdot\dots\cdot
T^{(\lambda)}_{a_{l}}
\label{eq13},
\end{equation}
where the $T_a^{(\lambda)}$ are irreducible representations of
(\ref{eq10}), having the property that certain polynomials, like
\be
C_1^{(\lambda)} : = T_aT_b g^{ab} = \lambda_1\cdot{\bf 1}
\label{eq14}
\ee

$(g^{ab}$ being the inverse of $g_{ab} : = {1\over 2} f_{ac}^d
f_{bd}^c$, the metric tensor on ${\cal G}$ ) are proportional
to {\bf 1}, and the coefficients $c^{a_1\dots a_l}$ apart from
(for definiteness) being totally antisymmetric, will consequently
be taken to satisfy additional requirements like $g_{ab}
c^{aba_3\dots a_l} \equiv 0, \dots$ (in accordance with the $r$
Casimir relations).
It is known (see e.g. \cite{Dixmier}) that $U_\lambda ({\cal G}
)$ decomposes, under the action of ${\cal G}$, into a direct sum of
finite dimensional irreducible ${\cal G}$-moduls,
\begin{equation}
U_{\lambda} = \oplus\sum_{(t)_{j}} U_\lambda^{(t)_{j}}
\label{eq15},
\end{equation}
$j=1\dots m(t)$, where each (tensor) representation $(t)$
occurs finitely many $(m_{(t)})$ times; the 1-dimensional
representation occurs only once and, as $[U_{\lambda}, U_\lambda]
= [{\cal G}_\lambda, U_\lambda]$ (see \cite{Warner}), it is
easy to see that $U_\lambda$ is also the direct sum
\begin{equation}
U_\lambda = {\bf C} \cdot {\bf 1} \oplus [U_\lambda, U_\lambda]
\label{eq16},
\end{equation}
implying that
\begin{equation}
TrT : = c_T
\label{eq17}
\end{equation}
defines an invariant trace on $U_\lambda$, $Tr[A,B] = 0$ -- which is
all one needs to conclude that (\ref{eq8}), with ${\cal A} = U_\lambda
({\cal G})$, will have infinitely many conserved quantities,
(\ref{eq7}). I am referring to the series of algebras $U_\lambda
({\cal G})$ as the `spherical series' as in the simplest case,
${\cal G} = SO(3)$, the (to be extended) map $\phi_{\hbar:
={1\over \lambda}}$,
\begin{eqnarray}
Y_{lm}(\theta,\phi) & = & \bigl (\sum c_{a_{1}\dots a_{l}}^{(lm)}
x_{a_{1}}\cdot\dots\cdot x_{a_{l}}
\bigr )_{{\vec x}= (\sin\theta \cos\phi, \sin \theta \sin\phi, 
\cos\theta)}\nonu\\
&\longrightarrow&\phi_\hbar(Y_{lm}) = \bar T^{(\lambda)}_{lm} :
= c_{(l)}^{(\lambda)}\sum c_{a_{1}\dots a_{l}}^{(lm)}
\bar T_{a_{1}\dots}^{(\lambda)} \bar T_{a_{l}}^{(\lambda)}
\label{eq18},
\end{eqnarray}
where $[\bar T_a^{(\lambda)}, \bar T_b^{(\lambda)}]={\epsilon_
{abc}\over \lambda} \bar T_c^{(\lambda)}, \sum^3_{a=1}
\bar T_a^{(\lambda)} \bar T_a^{(\lambda)}={\bf 1}$,
provides a one to one correspondence between $U_{\lambda}
(SO(3))$ and the (commutative, resp. Poisson-) algebra of
complex-valued functions on $S^2$ (cp. \cite{Hoppe8}, \cite{Hoppe14}).
Moreover, the overall normalisation may be chosen such that when
$\hbar={1\over\lambda}$ is appropriately taken to $0$,
\begin{eqnarray}
{1\over i\hbar}[\phi_\hbar(Y_{lm}),\phi_\hbar(Y_{l'm'})]
&\to & \{{Y_{lm}},{Y_{l' m'}}\}\nonu\\
&:=&{1\over sin\theta}\bigl({\partial Y_{lm}\over \partial\theta}
{\partial Y_{l'm'}\over\partial\phi}-(lm
\longleftrightarrow l'm')\bigr)
\label{eq19}
\end{eqnarray}
- -- which together with suitable properties under complex (hermitean)
conjugation, ... (see \cite{Bayen} for the definition of a star-product)
allows to call the associative multiplication in $U_\lambda(SO(3))$
a *-product on $S^2$:

\begin{equation}
Y_{lm}\ast Y_{l' m'}:=\phi^{-1}_\hbar(\phi_\hbar
(Y_{lm})\phi_\hbar(Y_{l' m '})\bigr)
\label{eq20}.
\end{equation}

For the `Torus-case', where a *-product may easily be written
directly in terms of functions on $T^2$,
\begin{equation}
f\ast g:= f\cdot g+\sum^{\infty}_{n=1}{(\lambda/2)^n\over n!}
\epsilon^{r_{1}s_{1}}\dots\epsilon^{r_{n}s_{n}}{\partial^n f
\over\partial\phi^{r_{1}}\dots\partial\phi^{r_{n}}}{\partial^n g
\over\partial\phi^{s_{1}}\dots\partial\phi^{s_{n}}},
\label{eqn21}
\end{equation}
the equation
\begin{equation}
\lambda\dot{x}_i=\epsilon_{ijk}x_j\ast x_k,
\label{eq22}
\end{equation}
viewed as an evolution-equation for a hypersurface in ${\bf R}^3$,
may then `at any stage' (s.b.) be compared with the time-harmonic
flow (\ref{eq9}) (which is equaivalent to (\ref{eq22}), $\lambda=0$).
In view of (\ref{eq9}) being equivalent to (cp. \cite{Hoppe})
\begin{equation}
\vec\nabla^2 t(x^1, x^2,x^3)=0
\label{eq23},
\end{equation}
it is tempting to interchange dependent and independent
variables also in (\ref{eq22}): making this `hodograph' transformation,
\begin{equation}
\phi^o=t, \phi^1, \phi^2\to x^i:=x^i(t,\phi^1,\phi^2)
\label{eq24},
\end{equation}
one first notes the purely `kinematical' consequences,
\begin{eqnarray}
\dot{\vec x}&\hat{=}& J(\vec\nabla\phi^1\times\vec\nabla\phi^2)\nonu\\
J:&=&\bigl|\bigl({\partial x^i\over \partial \phi^\alpha}\bigr)\bigr|
=\dot{\vec x} \bigl(\partial_1\vec x\times\partial_2\vec x\bigr)
=\bigl(\vec{\nabla}t\cdot(\vec\nabla\phi^1\times\vec\nabla\phi^2)
\bigr)^{-1}\nonu\\
\partial_t &\hat{=}&\dot{\vec x}\cdot\vec\nabla = J(\vec\nabla\phi^1
\times\vec\nabla\phi^2)\cdot\vec\nabla = :D_o\label{eq25}\\
{\partial\over \partial\phi^r} &\hat{=}&J(\vec\nabla t\times 
\vec\nabla\phi^r)\cdot\vec\nabla =:D^r\nonu\\
D^r\phi^s &=&\epsilon^{rs}, \quad[D^{\alpha}, D^\beta] = 0,
\nonu
\end{eqnarray}
$\alpha,\beta = 0,1,2$

while (\ref{eq22}) becomes
\begin{eqnarray}
\bigl(\vec\nabla\phi^1\times\vec\nabla\phi^2\bigr)_i
&=&{\epsilon_{ijk}\over \lambda J}
\quad e^{{\lambda\over 2}\epsilon_{rs}D^r\otimes D^s}
x_j\otimes x_k\Big|_{\rm Diag.}\nonu\\
&=&(\vec\nabla t)_i\nonu\\
&+&{1\over 2 J}\epsilon_{ijk}\sum^\infty_{l=1}
{(\lambda^2/4)^l\over (2l+1)!}(\epsilon_{rs}D^r
\otimes D^s)^{2l+1} x_j\otimes x_k\Big|_{\rm Diag.}\nonu\\
&=:&(\vec\nabla t)_i
+\sum^\infty_{l=1}(\lambda^2)^l
(\vec{F}_l(\vec\nabla t, \vec\nabla\phi^1,\vec\nabla\phi^2))_i
\label{eq26}.
\end{eqnarray}

Note that just as $\sum_i[x_i,\dot{x}_i]_\ast=0$ is a consequence
of (\ref{eq22}), solutions of (\ref{eq26}) will satisfy
\begin{eqnarray}
&\sum_i e^{{\lambda\over 2}\epsilon_{rs} D^r\otimes D^s} &
(x_i\otimes J(\vec\nabla\phi^1\times\vec\nabla\phi^2)_i -
J(\vec\nabla\phi^1\times\vec\nabla\phi^2)_i
\otimes x_i)\Big|_{\rm Diag.}\nonu\\
&& = 0
\label{eq27}
\end{eqnarray}
At least recursively, (\ref{eq26}) is still solvable, as expanding
the 3 unknown functions $t, \phi^1,\phi^2$ into powerseries in $\lambda^2$,
\begin{eqnarray}
t(\vec x) &=& T(\vec x)+\sum^\infty_{n=1}\lambda^{2n} t_n(\vec x)\nonu\\
\phi^r(\vec x) &=&{\bf\phi}^r(\vec x)+\sum^\infty_{n=1}\lambda^{2n}\phi^r_n
(\vec x)
\label{eq28},
\end{eqnarray}
the zero'th order (non-linear) ones \underline{are} solvable,
while all the higher ones are (recursively) linear; in particular,
all $t_n(\vec x)$ are given as solutions of Poisson's equation,
\begin{equation}
\vec\nabla^2 t_n(\vec x)=\vec\nabla G_n
\label{eq29},
\end{equation}
with $G_n$ only depending on the $t_{m<n}$, $\phi^1_{m<n}$ and
$\phi^2_{m<n}$. Of course, it would be desirable to derive
(from (\ref{eq26})) an equation only involving $t$.

{\bf Acknowledgement:}
I would like to thank M. Bordemann and M. Seifert for valuable
discussions, the organizers and participants of `Deformation
theory, symplectic geometry and applications' for the
stimulating atmosphere at the Monte Verita, and the 
Eidgen\"ossische Technische Hochschule Z\"urich for
hospitality.

\end{document}